\documentclass[11pt,letterpaper]{article}
\usepackage{amssymb}
\usepackage{epsfig}
\usepackage{cite}
\usepackage[headsep=0pt, head=30pt,foot=30pt,includeheadfoot,dvips,
  inner=3cm,outer=3cm,top=2cm, bottom=2cm]{geometry}

\newcommand{\labe}[1]{\label{equ:#1}}
\newcommand{\labf}[1]{\label{fig:#1}}
\newcommand{\labt}[1]{\label{tab:#1}}
\newcommand{\refe}[1]{\ref{equ:#1}}
\newcommand{\reff}[1]{\ref{fig:#1}}
\newcommand{\reft}[1]{\ref{tab:#1}}
\newcommand{\Ref}[1]{Ref.~\cite{#1}}
\newcommand{\Eq}[1]{Eq.~(\refe{#1})}
\newcommand{\Eqss}[2]{Eqs.~(\refe{#1}) and (\refe{#2})}
\newcommand{\Eqssor}[2]{Eqs.~(\refe{#1}) or (\refe{#2})}
\newcommand{\Eqsss}[3]{Eqs.~(\refe{#1}), (\refe{#2}) and (\refe{#3})}
\newcommand{\Fig}[1]{Fig.~\reff{#1}}
\newcommand{\Table}[1]{Table~\reft{#1}}

\def\collab#1{{\bf #1\rm}}
\def\etalcollab#1{\etal\,(\collab{#1}),} 

\def\Bd{$B{^0_d}$}
\def\Bdbar{$\overline B{^0_d}$}
\def\Bs{$B{^0_s}$}
\def\Bsbar{$\overline B{^0_s}$}
\def\Bqq{$B{^0_q}$}
\def\Bqbar{$\overline B{^0_q}$}
\def\BdBdbar{\Bd--\Bdbar}
\def\BsBsbar{\Bs--\Bsbar}
\def\BqBqbar{\Bqq--\Bqbar}
\def\BBbar{$B{^0}\hbox{--}\overline B{^0}$}
\def\KKbar{$K{^0}\hbox{--}\overline K{^0}$}

\def\EQN#1{\labe{#1}}
\let\DELTA=\Delta 
\def\IndexPageno#1{}
\def\lsim{\,\hbox{\char'056}\,} 
\def\etal{\hbox{\it et~al.}} 
\def\arns#1,#2(#3)
   {{\rm Ann.\,Rev.\,Nucl.\,Sci.\ }{\bf #1}, {\rm#2} {\rm(#3)}} 
\def\epjC#1,#2(#3){{\rm Eur.\,Phys.\,J.\,}{\bf C#1}, {\rm#2} {\rm(#3)}} 
\def\npB#1,#2(#3){{\rm Nucl.\,Phys.\,}{\bf B#1}, {\rm#2} {\rm(#3)}} 
\def\ptp#1,#2(#3){{\rm Prog.\,Theor.\,Phys.\,}{\bf #1}, {\rm#2} {\rm(#3)}} 
\def\plB#1,#2(#3){{\rm Phys.\,Lett.\,}{\bf B#1}, {\rm#2} {\rm(#3)}} 
\def\nim#1,#2(#3)
   {{\rm Nucl.\,Instrum.\,Methods\,}{\bf #1}, {\rm#2} {\rm(#3)}}
\def\prl#1,#2(#3){{\rm Phys.\,Rev.\,Lett.\,}{\bf #1}, {\rm#2} {\rm(#3)}} 
\def\prD#1,#2(#3){{\rm Phys.\,Rev.\,}{\bf D#1}, {\rm#2} {\rm(#3)}} 
\def\zpC#1,#2(#3){{\rm Z.\,Phys.\,}{\bf C#1}, {\rm#2} {\rm(#3)}} 
\def\ijmpA#1,#2(#3)
   {{\rm Int.\,J.\,Mod.\,Phys.\,}{\bf A#1}, {\rm#2} {\rm(#3)}}
\def\npBps#1,#2(#3){{\rm Nucl.\,Phys.\,(Proc.\,Supp.),}{\bf B#1},
{\rm#2} {\rm(#3)}} 
\def\reference#1{\bibitem{#1}}

\def\lsim{\mathrel{\rlap{\lower4pt\hbox{\hskip1pt$\sim$}}
    \raise1pt\hbox{$<$}}}                

\begin{document}
\pagestyle{empty}
\begin{flushright}
LPHE 2006-013 \\
May 3, 2006 \\~
\end{flushright}

\vfill

\begin{center}

{\LARGE\bf \boldmath $B{^0}\hbox{--}\overline B{^0}$ MIXING}
\\[6ex] {\Large O.~SCHNEIDER} \\[1ex]
{\it Laboratoire de Physique des Hautes Energies} \\
{\it Ecole Polytechnique F\'ed\'erale de Lausanne (EPFL) \\
CH--1015 Lausanne, Switzerland} \\[1ex]
{\it e-mail:} {\tt Olivier.Schneider@epfl.ch} \\ ~

\vfill

\begin{minipage}{0.8\textwidth}
The subject of particle-antiparticle mixing in the neutral $B$ meson systems 
is reviewed. The formalism of $B{^0}\hbox{--}\overline B{^0}$ mixing 
is recalled and 
basic Standard Model predictions are given, before experimental issues are 
discussed and the latest combinations of experimental results on 
mixing parameters are presented, including those on 
mixing-induced $CP$ violation, mass differences, and decay-width differences.
Finally, time-integrated mixing results are used to improve our knowledge on 
the fractions of the various $b$-hadron species produced in $Z$ decays and 
at high-energy colliders.
\end{minipage}

\vfill ~ \vfill

{\it To appear in the 2006 edition of the 
``Review of Particle Physics'',
} \\ {\it  
W.-M.~Yao et al.\ (Particle Data Group), 
J.\ Phys.\ G: Nucl.\ Part.\ Phys.\ 33, 1 (2006).
}\\[2ex]
\end{center}

\newpage
~

\newpage
\pagestyle{plain}\setcounter{page}{1}   


\begin{center}
{\LARGE\bf \boldmath \BBbar\ MIXING}

\vspace{3mm}
{\em
Updated April 2006
by O.\ Schneider (Ecole Polytechnique F\'ed\'erale de Lausanne)
}
\vspace{5mm}
\end{center}

There are two neutral \BBbar\ meson systems, \BdBdbar\ and \BsBsbar\
(generically denoted \BqBqbar, $q=s,d$), which 
exhibit particle-antiparticle mixing\cite{textbooks}.
This mixing phenomenon is described in \Ref{CP_review}.
In the following, we adopt the notation introduced in \Ref{CP_review},
and assume $CPT$ conservation throughout.
In each system, the light (L) and heavy (H) mass eigenstates,
\begin{equation} 
|B_{\rm L,H}\rangle = p | B{^0_q}\rangle \pm q |\overline B{^0_q}\rangle \,,
\EQN{eigenstates}
\end{equation} 
have a mass difference $\DELTA m_q = m_{\rm H} -m_{\rm L} > 0$,
and a total decay width difference
$\DELTA \Gamma_q = \Gamma_{\rm L} -\Gamma_{\rm H}$.
In the absence of $CP$ violation in the mixing,
$|q/p|=1$, these differences are given by $\DELTA m_q =2|M_{12}|$
and $|\DELTA \Gamma_q| =2|\Gamma_{12}|$, where $M_{12}$ and $\Gamma_{12}$
are the off-diagonal elements of the mass and decay matrices\cite{CP_review}.
The evolution of a pure $| B{^0_q}\rangle$ or
$|\overline B{^0_q}\rangle$ state at $t=0$ is given by
\begin{eqnarray} 
| B{^0_q}(t)\rangle &=& g_+(t) \,| B{^0_q}\rangle
                     + \frac{q}{p} \, g_-(t) \,|\overline B{^0_q}\rangle \,,
\EQN{time_evol1} 
\\
|\overline B{^0_q}(t)\rangle &=& g_+(t) \,|\overline B{^0_q}\rangle
                     + \frac{p}{q} g_-(t) \,| B{^0_q}\rangle \,,
\EQN{time_evol2} 
\end{eqnarray} 
which means that the flavor states remain unchanged ($+$) or oscillate
into each other ($-$) with time-dependent probabilities proportional to
\begin{equation} 
\left| g_{\pm}(t)\right|^2 = \frac{e^{-\Gamma_q t}}{2}
\left[ \cosh\!\left(
\frac{\DELTA\Gamma_q}{2}\,t\right) \pm \cos(\DELTA m_q\,t)\right] \,, 
\EQN{cosh_cos}
\end{equation} 
where $\Gamma_q = (\Gamma_{\rm H} +\Gamma_{\rm L})/2$.
In the absence of $CP$ violation, the time-integrated mixing probability
$\int \left| g_-(t)\right|^2 dt /
(\int \left| g_-(t)\right|^2 dt + \int \left| g_+(t)\right|^2 dt)$
is given by 
\begin{equation} 
\chi_q = \frac{x_q^2+y_q^2}{2(x_q^2+1)} \,, ~~~{\rm where}~~~
x_q = \frac{\DELTA m_q}{\Gamma_q}  
\,, ~~~
y_q = \frac{\DELTA \Gamma_q}{2\Gamma_q} \,.
\EQN{chi}
\end{equation} 

\section*{Standard Model predictions and phenomenology}

In the Standard Model, the transitions \Bqq$\to$\Bqbar\ and \Bqbar$\to$\Bqq\ 
are due to the weak interaction.
They are described, 
at the lowest order, by box diagrams involving
two $W$~bosons and two up-type quarks (see \Fig{box}), 
as is the case for \KKbar\ mixing.
However, the long range 
interactions arising from intermediate virtual states are negligible 
for the neutral $B$ meson systems,
because the large $B$ mass is off the region of hadronic resonances. 
The calculation of the dispersive and 
absorptive parts of the box diagrams yields the following predictions 
for the off-diagonal element of the mass and decay matrices\cite{Buras84},
\begin{figure}\begin{center}
~
\hfill
\epsfig{figure=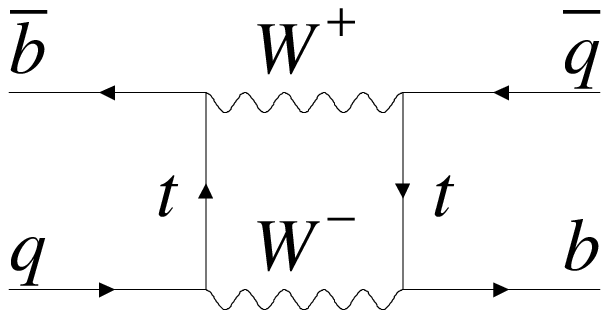,width=0.4\textwidth,clip=t}%
\hfill
\epsfig{figure=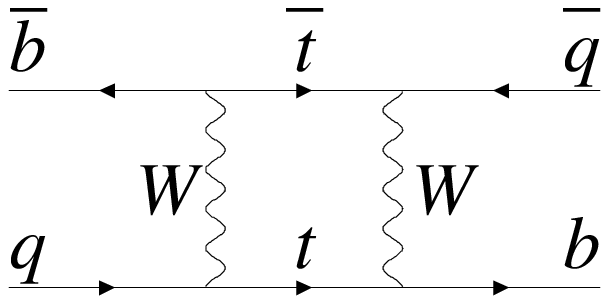,width=0.4\textwidth,clip=t}%
\hfill
~
\caption{
Dominant box diagrams for the 
\Bqq$\to$\Bqbar\ transitions ($q = d$ or $s$). Similar 
diagrams exist where one or both $t$ quarks are 
replaced with $c$ or $u$ quarks.
} 
\labf{box}
\end{center}\end{figure}
\begin{eqnarray} 
\hspace{-2.5ex} M_{12} \hspace{-1ex}&=&\hspace{-1ex} - \frac{
           G_F^2 m_W^2 \eta_B m_{B_q} B_{B_q} f_{B_q}^2}{12\pi^2} 
           \, S_0(m_t^2/m_W^2) \, (V_{tq}^* V_{tb}^{})^2 \,,
\EQN{M_12} \\ 
\hspace{-2.5ex} \Gamma_{12} \hspace{-1ex}&=&\hspace{-1ex} \frac{
           G_F^2 m_b^2 \eta'_B m_{B_q} B_{B_q} f_{B_q}^2}{8\pi} 
\hspace{-0.5ex}
           \left[ (V_{tq}^* V_{tb}^{})^2 + 
                   V_{tq}^* V_{tb}^{} V_{cq}^* V_{cb}^{} \,\, 
                         \hspace{-0.5ex}{\cal O}\!\left(\frac{m_c^2}{m_b^2}\right) 
                 + (V_{cq}^* V_{cb}^{})^2 \,\,
                         \hspace{-0.5ex}{\cal O}\!\left(\frac{m_c^4}{m_b^4}\right) 
           \right] \hspace{-0.5ex}\,,
\end{eqnarray} 
\noindent 
where $G_F$ is the Fermi constant, $m_W$ the $W$ boson mass,
and $m_i$ the mass of quark $i$;
$m_{B_q}$, $f_{B_q}$ and $B_{B_q}$ are the \Bqq\ mass, 
weak decay constant and bag parameter, respectively.
The known function $S_0(x_t)$ 
can be approximated very well by
$0.784\,x_t^{0.76}$\cite{Buras_Fleischer_HeavyFlavorsII},
and $V_{ij}$ are the elements of the CKM matrix\cite{CKM}.
The QCD corrections $\eta_B$ and $\eta'_B$ are of order unity. 
The only non-negligible contributions to $M_{12}$ are from box diagrams
involving two top quarks. 
The phases of $M_{12}$ and $\Gamma_{12}$ satisfy
\begin{equation} 
\phi_{M} - \phi_{\Gamma} = 
\pi + {\cal O}\left(\frac{m^2_c}{m^2_b} \right) \,,
\EQN{phasediff}
\end{equation} 
implying that the mass eigenstates  
have mass and width differences of opposite signs. This means that,
like in the $K{^0}\hbox{--}\overline K{^0}$ system, the 
heavy state is expected to have a smaller decay width than 
that of the light state: 
$\Gamma_{\rm H} < \Gamma_{\rm L}$.
Hence, $\DELTA\Gamma = \Gamma_{\rm L} -\Gamma_{\rm H}$
is expected to be positive in the Standard Model.

Furthermore, the quantity 
\begin{equation} 
\left|\frac{\Gamma_{12}}{M_{12}}\right| \simeq \frac{3\pi}{2}
\frac{m^2_b}{m^2_W} \frac{1}{S_0(m_t^2/m_W^2)} 
\sim {\cal O}\left(\frac{m^2_b}{m^2_t} \right)
\EQN{G12overM12} 
\end{equation} 
is small, and a power expansion of $|q/p|^2$ yields
\begin{equation} 
\left|\frac{q}{p}\right|^2 = 1 + \left|\frac{\Gamma_{12}}{M_{12}}\right| 
\sin(\phi_{M}-\phi_{\Gamma})
+ {\cal O}\left( \left|\frac{\Gamma_{12}}{M_{12}}\right|^2\right) \,.
\end{equation} 
Therefore, considering both 
\Eqss{phasediff}{G12overM12},
the $CP$-violating parameter
\begin{equation} 
1 - \left|\frac{q}{p}\right|^2 \simeq 
{\rm Im}\left(\frac{\Gamma_{12}}{M_{12}}\right) 
\end{equation} 
is expected to be very small: $\sim {\cal O}(10^{-3})$ for the 
\BdBdbar\ system and $\lsim {\cal O}(10^{-4})$ 
for the \BsBsbar\ system\cite{Bigi}.

In the approximation of negligible $CP$ violation in mixing, 
the ratio $\DELTA\Gamma_q/\DELTA m_q$ is equal to the 
small quantity $\left|\Gamma_{12}/M_{12}\right|$ of \Eq{G12overM12}; it is
hence independent of CKM matrix elements, {\it i.e.},
the same for the \BdBdbar\ and \BsBsbar\ systems. It can be calculated 
with lattice QCD techniques; typical results are $\sim 5 \times 10^{-3}$
with quoted uncertainties of $\sim 30\%$.
Given the current experimental knowledge  
on the mixing parameter $x_q$ 
(obtained from published results only), 
\begin{equation} 
\cases{
x_d = 0.776 \pm 0.008 & (\BdBdbar\ system) \cr
x_s > 19.9 ~ \hbox{at 95\%~CL} & (\BsBsbar\ system) \cr}
 \,,
\end{equation} 
the Standard Model thus predicts that $\DELTA\Gamma_d/\Gamma_d$ is very 
small (below 1\%), 
but $\DELTA\Gamma_s/\Gamma_s$ considerably larger 
($\sim 10\%$). These width differences are
caused by the existence of final states to which 
both the \Bqq\ and \Bqbar\ mesons can decay. Such decays involve 
$b \to c\overline{c}q$ quark-level transitions,
which are 
Cabibbo-suppressed if $q=d$ and Cabibbo-allowed if $q=s$.

\section*{Experimental issues and methods for oscillation analyses}

Time-integrated measurements of \BBbar\ mixing
were published for the first time in 1987 by UA1\cite{UA1_CP} 
and ARGUS\cite{ARGUS_CP}, 
and since then by many other experiments.
These measurements are typically based on counting same-sign and opposite-sign 
lepton pairs from the semileptonic decay of the produced $b\overline{b}$ pairs.
Such analyses cannot easily separate the contributions from the 
different $b$-hadron species, therefore, the clean environment 
of $\Upsilon(4S)$ machines (where only \Bd\ and charged $B_u$ mesons 
are produced) is in principle best suited to measure $\chi_d$.

However, better sensitivity is obtained from time-dependent analyses
aiming at the direct measurement of the oscillation frequencies 
$\DELTA m_d$ and $\DELTA m_s$,
from the proper time
distributions of \Bd\ or \Bs\ candidates 
identified through their decay in (mostly) flavor-specific modes, and
suitably tagged as mixed or unmixed.
This is particularly true for the \BsBsbar\ 
system, where the large 
value of $x_s$ implies maximal mixing, {\it i.e.}, $\chi_s \simeq 1/2$.
In such analyses, the \Bd\ or \Bs\ mesons are 
either fully reconstructed, partially reconstructed from a charm meson, 
selected from a lepton 
with the characteristics of a $b\to\ell^-$ decay,
or selected from a reconstructed displaced vertex. 
At high-energy colliders (LEP, SLC, Tevatron), 
the proper time $t=\frac{m_B}{p}L$ is measured 
from the distance $L$ between the production vertex and 
the $B$ decay vertex, 
and from an estimate of the $B$ momentum $p$.
At asymmetric $B$ factories (KEKB, PEP-II), producing 
$e^+e^-\to\Upsilon(4S) \to\hbox{\Bd\Bdbar}$ events with a boost
$\beta\gamma$ ($=0.425$, $0.55$),
the proper time difference between the two $B$ candidates
is estimated as $\DELTA t \simeq \frac{\DELTA z}{\beta\gamma c}$, 
where $\DELTA z$ is the spatial separation between the 
two $B$ decay vertices along the boost direction. 
In all cases, the good resolution needed on the vertex positions 
is obtained with silicon detectors. 

The average statistical significance ${\cal S}$ 
of a \Bd\ or \Bs\ oscillation signal can be approximated as\cite{amplitude}
\IndexPageno{Bstatsigm}
\begin{equation} 
{\cal S} \approx \sqrt{N/2} \,f_{\rm sig}\, (1-2\eta)\,
e^{-(\DELTA m\,\sigma_t)^2/2}  \,, 
\EQN{significance} 
\end{equation} 
where $N$ is the number of selected and tagged candidates, 
$f_{\rm sig}$ is the fraction of signal in that sample, 
$\eta$ is the total mistag probability, 
and $\sigma_t$ is the resolution on proper time (or proper time difference). 
The quantity ${\cal S}$ decreases very quickly as 
$\DELTA m$ increases; this dependence is controlled by $\sigma_t$,
which is therefore a critical parameter for $\DELTA m_s$ analyses. 
At high-energy colliders, the proper time resolution 
$\sigma_t \sim \frac{m_B}{\langle p\rangle} \sigma_L 
\oplus t \frac{\sigma_p}{p}$ 
includes a constant contribution due to the decay length resolution 
$\sigma_L$ (typically 0.05--0.3~ps), and a term due to the 
relative momentum resolution $\sigma_p/p$ (typically 10--20\%
for partially reconstructed decays), 
which increases with proper time.
At $B$ factories,
the boost of the $B$ mesons is estimated from the known beam energies,
and the term due to the spatial resolution dominates
(typically 1--1.5~ps because of the much smaller $B$ boost).

In order to tag a $B$ candidate 
as mixed or unmixed, it is necessary
to determine its flavor 
both in the initial state and in the final state. 
The initial and final state mistag probabilities,\IndexPageno{Bmistagm}
$\eta_i$ and $\eta_f$, degrade ${\cal S}$
by a total factor $(1-2\eta)=(1-2\eta_i)(1-2\eta_f)$.
In lepton-based analyses, the final state is tagged by the charge of 
the lepton from $b\to\ell^-$ decays; the largest contribution to $\eta_f$ 
is then due to $\overline{b}\to\overline{c}\to\ell^-$ decays. 
Alternatively, the charge of a 
reconstructed charm meson ($D^{*-}$ from \Bd\ or $D_s^-$ from \Bs), 
or that of a kaon hypothesized to come from a $b\to c\to s$
decay\cite{SLD_dmd_prelim}, can be used.
For fully inclusive analyses based on topological 
vertexing, final state tagging techniques include 
jet charge\cite{ALEPH_dmd_prelim} and charge 
dipole\cite{SLD_dms_dipole,DELPHI_dmd_dms} methods.

At high-energy colliders, the methods to tag the initial state 
({\it i.e.}, the state at production), 
can be divided into two groups: the ones 
that tag the initial charge of the $\overline{b}$ quark contained in the 
$B$ candidate itself (same-side tag),\IndexPageno{bmixam}
 and the ones that tag the initial 
charge of the other $b$ quark produced in the event (opposite-side tag). 
On the same side, the charge of a track from the primary vertex is 
correlated with the production state of the $B$ if that track is a decay
product of a $B^{**}$ state or the first particle in the fragmentation 
chain\cite{CDF_dmd,ALEPH_dms}.
Jet- and vertex-charge techniques work on both sides and on the opposite
side, respectively. 
Finally, the charge of a lepton from $b\to\ell^-$ or of a kaon
from $b\to c\to s$ can be used as opposite side tags,
keeping in mind that their performance is degraded due to integrated mixing.
At SLC, the beam polarization produced a sizeable forward-backward 
asymmetry in the $Z\to b\overline{b}$ decays, and provided another
very interesting and effective initial state tag based on the polar angle 
of the $B$ candidate\cite{SLD_dms_dipole}.
Initial state tags have also been combined to 
reach $\eta_i \sim 26\%$ at LEP\cite{ALEPH_dms,DELPHI_dms_dgs},
or even 22\% at SLD\cite{SLD_dms_dipole} with full efficiency. 
In the case $\eta_f=0$, this corresponds to an effective tagging efficiency
$Q=\epsilon D^2=\epsilon(1-2\eta)^2$, 
where $\epsilon$ is the tagging efficiency,
in the range $23-31\%$. 
The equivalent figure achieved by CDF 
during Tevatron Run~I was $\sim3.5\%$\cite{Paulini}
reflecting the fact that tagging is more difficult at hadron colliders.
The current CDF and D\O\ analyses of Tevatron Run~II data reach 
$\epsilon D^2 = (1.5\pm0.1)\%$\cite{CDF2_dms_prelim} 
and $(2.5\pm0.2)\%$\cite{DZERO_dms} for opposite-side tagging, 
while same-side kaon tagging (for \Bs\ oscillation analyses) 
is contributing an additional $(3.4\pm1.0)\%$ 
at CDF\cite{CDF2_dms_prelim}.

At $B$ factories, the flavor of a \Bd\ meson at production cannot 
be determined, since the two neutral $B$ mesons produced in a 
$\Upsilon(4S)$ decay evolve in a coherent $P$-wave state where they 
keep opposite flavors at any time.
However, as soon as one of them decays, the other follows a time-evolution
given by \Eqssor{time_evol1}{time_evol2},
where $t$ is replaced with $\DELTA t$
(which will take negative values half of the time).
Hence, the ``initial state''
tag of a $B$ can be taken as the final state tag of the other $B$. 
Effective tagging efficiencies $Q$ of 30\%
are achieved by BABAR 
and Belle\cite{BABAR_Belle_tagging}, 
using different techniques including $b \to \ell^-$ and 
$b\to c\to s$ tags. 
It is worth noting that, in this case, 
mixing of the other $B$ ({\it i.e.}, the coherent mixing occurring before
the first $B$ decay) does not contribute to the mistag probability.

In the absence of experimental observation of a decay-width difference,
oscillation analyses typically neglect $\DELTA\Gamma$ in \Eq{cosh_cos},
and describe the data with the physics functions
$\Gamma e^{-\Gamma t} (1 \pm \cos( \DELTA m t))/2$
(high-energy colliders) or
$\Gamma e^{-\Gamma |\DELTA t|} (1 \pm \cos( \DELTA m \DELTA t))/4$
(asymmetric $\Upsilon(4S)$ machines).
As can be seen from \Eq{cosh_cos}, a non-zero value of $\DELTA\Gamma$
would effectively reduce the oscillation amplitude with a small 
time-dependent factor that would be very difficult to distinguish 
from time resolution effects. 
Measurements of $\DELTA m_d$ are usually extracted from the data
using a maximum likelihood fit.
To extract information useful for the interpretation of \Bs\ oscillation searches 
and for the combination of their results, 
a method\cite{amplitude} is followed
in which a \Bs\ oscillation amplitude ${\cal A}$
is measured as a function of a fixed test value of $\DELTA m_s$, 
using a maximum likelihood fit based on the functions
$\Gamma_s e^{-\Gamma_s t} (1 \pm {\cal A} \cos( \DELTA m_s t))/2$. 
To a good approximation, the statistical uncertainty on ${\cal A}$
is Gaussian and equal to $1/{\cal S}$ from \Eq{significance}.
If $\DELTA m_s$ is equal to its true value, one expects
${\cal A} = 1 $ within the total uncertainty $\sigma_{\cal A}$;
in case a signal is seen, its observed (or expected) 
significance will be defined as 
${\cal A}/\sigma_{\cal A}$ (or $1/\sigma_{\cal A}$).
However, if $\DELTA m_s$ is (far) below its
true value, a measurement consistent with ${\cal A} = 0$ is expected.
A value of $\DELTA m_s$ can be excluded at 95\%~CL
if ${\cal A} + 1.645\,\sigma_{\cal A} \le 1$ (since the integral
of a normal distribution from $-\infty$ to $1.645$ is equal to 0.95). 
Because of the proper time resolution, the quantity $\sigma_{\cal A}(\DELTA m_s)$
is a steadily increasing function of $\DELTA m_s$. 
We define the sensitivity for 95\%~CL exclusion of $\DELTA m_s$ values
(or for a $3\,\sigma$ or $5\,\sigma$ observation of \Bs\ oscillations) 
as the value of $\DELTA m_s$ for which 
$1/\sigma_{\cal A}=1.645$ (or $1/\sigma_{\cal A}=3$ or $5$).

\section*{\boldmath \Bd\ mixing studies}


Many \BdBdbar\ oscillations analyses have been 
published\cite{WARNING} by the ALEPH\cite{ALEPH_dmd}, 
BABAR\cite{BABAR_dmd}, Belle\cite{Belle_dmd},
CDF\cite{CDF_dmd}, 
DELPHI\cite{DELPHI_dmd_dms,DELPHI_dmd}, 
L3\cite{L3_dmd}, and OPAL\cite{OPAL_dmd} collaborations.
Although a variety of different techniques have been used, the 
individual $\DELTA m_d$ 
results obtained at high-energy colliders have remarkably similar precision.
Their average is compatible with the recent and more precise measurements 
from asymmetric $B$ factories.
The systematic uncertainties are not negligible; 
they are often dominated by sample composition, mistag probability,
or $b$-hadron lifetime contributions.
Before being combined, the measurements are adjusted on the basis of a 
common set of input values, including the $b$-hadron lifetimes and fractions
published in this {\it Review}. Some measurements are statistically correlated. 
Systematic correlations arise both from common physics sources (fragmentation 
fractions, lifetimes, branching ratios of $b$~hadrons), and from purely 
experimental or algorithmic effects (efficiency, resolution, tagging, 
background description). Combining all published measurements%
\cite{DELPHI_dmd_dms,CDF_dmd,ALEPH_dmd,BABAR_dmd,Belle_dmd,DELPHI_dmd,L3_dmd,OPAL_dmd}
and accounting for all identified correlations 
yields 
$\DELTA m_d = {\rm 0.507 \pm 0.003 (stat) \pm 0.003 (syst)}~\hbox{ps}^{-1}$\cite{HFAG}, 
a result now dominated by the $B$ factories.

On the other hand, ARGUS and CLEO have published time-integrated 
measurements\cite{ARGUS_chid,CLEO_chid_CP,CLEO_chid_CP_y}, 
which average to $\chi_d = 0.182 \pm 0.015$.
Following \Ref{CLEO_chid_CP_y}, 
the width difference $\DELTA \Gamma_d$ could 
in principle be extracted from the
measured value of $\Gamma_d$ and the above averages for 
$\DELTA m_d$ and $\chi_d$ 
(see \Eq{chi}),
provided that $\DELTA \Gamma_d$ has a negligible impact on 
the $\DELTA m_d$ measurements.
However, direct time-dependent studies published  
by DELPHI\cite{DELPHI_dmd_dms} and 
BABAR\cite{BABAR_DGd_qp} yield stronger constraints, which can be combined 
to yield 
$\rm{sign}(\rm{Re} \lambda_{CP}) \DELTA \Gamma_d/\Gamma_d 
= 0.009 \pm 0.037$\cite{HFAG}.

Assuming $\DELTA \Gamma_d =0$ and no $CP$ violation in mixing, 
and using the measured \Bd\ lifetime of $1.530\pm0.009~{\rm ps}^{-1}$,
the $\DELTA m_d$ and $\chi_d$ results are combined to yield the 
world average
\begin{equation} 
\DELTA m_d = 0.507 \pm 0.005~\hbox{ps}^{-1} 
\EQN{dmdw}
\end{equation} 
or, equivalently,
\begin{equation} 
\chi_d=0.188\pm 0.003\,.  
\EQN{chidw}
\end{equation} 

Evidence for $CP$ violation in \Bd\ mixing has been searched for,
both with flavor-specific and inclusive \Bd\ decays, 
in samples where the initial 
flavor state is tagged. In the case of semileptonic 
(or other flavor-specific) decays, 
where the final state tag is 
also available, the following asymmetry\cite{CP_review}
\begin{equation} 
 {\cal A}_{\rm SL} = 
\frac{
N(\hbox{\Bdbar}(t) \to \ell^+           \nu_{\ell} X) -
N(\hbox{\Bd}(t)    \to \ell^- \overline{\nu}_{\ell} X) }{
N(\hbox{\Bdbar}(t) \to \ell^+           \nu_{\ell} X) +
N(\hbox{\Bd}(t)    \to \ell^- \overline{\nu}_{\ell} X) } 
\simeq 1 - |q/p|^2_d 
\end{equation} 
has been measured, either in time-integrated analyses at 
CLEO
\cite{CLEO_chid_CP_y,CLEO_CP_semi},
CDF\cite{CDF_CP_semi} and D\O\cite{DZERO_Asl},
or in time-dependent analyses at 
LEP\cite{OPAL_CP_semi,DELPHI_CP,ALEPH_CP} and 
BABAR\cite{BABAR_DGd_qp,BABAR_CP_semi} and Belle\cite{Belle_Asl}.
In the inclusive case, also investigated at 
LEP\cite{DELPHI_CP,ALEPH_CP,OPAL_CP_incl},
no final state tag is used, and the asymmetry\cite{incl_asym}
\begin{equation} 
\frac{
N(\hbox{\Bd}(t) \to {\rm all}) -
N(\hbox{\Bdbar}(t) \to {\rm all}) }{
N(\hbox{\Bd}(t) \to {\rm all}) +
N(\hbox{\Bdbar}(t) \to {\rm all}) } 
\simeq
{\cal A}_{\rm SL} \left[ \frac{x_d}{2} \sin(\DELTA m_d \,t) - 
\sin^2\left(\frac{\DELTA m_d \,t}{2}\right)\right] 
\end{equation} 
must be measured as a function of the proper time to extract information 
on $CP$ violation.
In all cases, asymmetries compatible with zero have been found,  
with a precision limited by the available statistics. A simple 
average of all published 
results for the \Bd\ meson%
\cite{CLEO_chid_CP_y,BABAR_DGd_qp,CLEO_CP_semi,OPAL_CP_semi,ALEPH_CP,BABAR_CP_semi,OPAL_CP_incl}
yields
${\cal A}_{\rm SL} = {\rm -0.005 \pm 0.012}$,
or 
$|q/p|_d = 1.0026 \pm 0.0059$,
a result which does not yet constrain the Standard Model.

The $\DELTA m_d$ result of \Eq{dmdw} provides an estimate of $2|M_{12}|$, 
and can be used, 
together with \Eq{M_12}, 
to extract the magnitude of the CKM matrix element $V_{td}$ 
within the Standard Model\cite{CKM_review}. 
%
%
%
%
%
%
%
%
The main experimental 
uncertainties on the resulting estimate of $|V_{td}|$ come from 
$m_t$ and $\DELTA m_d$; however, the extraction is at present 
completely dominated by the uncertainty on the hadronic 
matrix element $f_{B_d} \sqrt{B_{B_d}} = 244\pm26$~MeV 
obtained from lattice QCD calculations\cite{lattice_QCD}.

\section*{\boldmath \Bs\ mixing studies}
\IndexPageno{Bsmixingm}

\BsBsbar\ oscillations have been the subject of many 
studies from ALEPH\cite{ALEPH_dms_final}, 
DELPHI\cite{DELPHI_dmd_dms,DELPHI_dms_dgs,DELPHI_dms}, 
OPAL\cite{OPAL_dms}, SLD\cite{SLD_dms_dipole,SLD_dms_ds,SLD_dms_prelim}, 
CDF\cite{CDF2_dms_prelim,CDF1_dms} and D\O\cite{DZERO_dms,DZERO_dms_prelim}. 
The most sensitive analyses at LEP appear to be the ones based 
on inclusive lepton samples. Because of their better 
proper time resolution, the small data samples analyzed 
inclusively at SLD, as well as the fully reconstructed $B_s$ decays 
at LEP and at the Tevatron, are also very useful 
to explore the high $\DELTA m_s$ region.

All results are limited by the available statistics.
They can easily be combined, since all experiments provide 
measurements of the \Bs\ oscillation amplitude. 
All published results%
\cite{SLD_dms_dipole,DELPHI_dmd_dms,DELPHI_dms_dgs,ALEPH_dms_final,DELPHI_dms,OPAL_dms,SLD_dms_ds,CDF1_dms}
are averaged\cite{HFAG} to yield the combined amplitudes 
${\cal A}$ shown in \Fig{amplitude} (top) as a function of $\DELTA m_s$. 
The individual results 
have been adjusted to common physics inputs, and all known correlations 
have been accounted for; 
the sensitivities of the inclusive analyses, 
which depend directly through \Eq{significance} 
on the assumed fraction $f_s$
of \Bs\ mesons in an unbiased sample of weakly-decaying $b$~hadrons, 
have also been rescaled to a common
average of 
$f_s = 0.102 \pm 0.009$. 
The combined sensitivity for 95\%~CL exclusion of $\DELTA m_s$ values is found to be
       18.2~ps$^{-1}$. 
All values of $\DELTA m_s$ below 
       14.4~ps$^{-1}$ 
       are excluded at 95\%~CL, which we express as
\begin{equation} 
\DELTA m_s > 14.4~\hbox{ps}^{-1} ~~~ \hbox{at 95\%~CL} \,.
\EQN{dms}
\end{equation} 
The values between 
       14.4 and 21.8~ps$^{-1}$ 
cannot be excluded, because 
the data is compatible with a signal in this region. However,
the largest deviation from ${\cal A}=0$ in this range is a 1.9\,$\sigma$ effect
only, so no signal can be claimed. 

\begin{figure}
\vspace{-25mm}
\begin{center}
\epsfig{figure=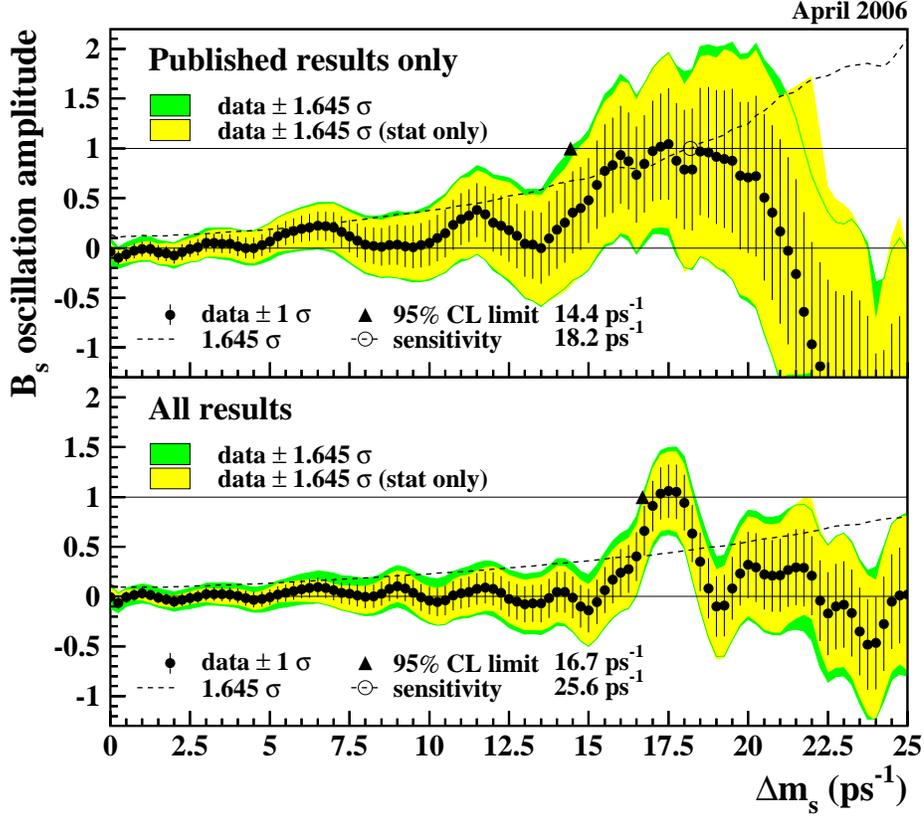,width=0.85\textwidth}
\caption{
Combined measurements of the \Bs\ oscillation amplitude as a 
function of $\DELTA m_s$, based on published results only (top) or 
on all published and unpublished results (bottom) 
available at the end of April 2006. 
The measurements are dominated by statistical uncertainties. 
Neighboring points are statistically correlated.
} 
\labf{amplitude}
\end{center}\end{figure}

The above average does not include the very recent 
results from Tevatron Run~II, based on 1~fb$^{-1}$ of data.
In a paper submitted for publication\cite{DZERO_dms}, 
D\O\ reports the first direct two-sided bound established by a single 
experiment of $17 < \DELTA m_s < 21~{\rm ps}^{-1}$ (90\%~CL) 
and a most probable value of 19~ps$^{-1}$ with an observed (expected) 
significance of 2.5\,$\sigma$ (0.9\,$\sigma$).
A preliminary and subsequent analysis from CDF\cite{CDF2_dms_prelim} is 
more sensitive and leads to the first direct evidence of \Bs\ oscillations 
and the following measurement: 
\begin{equation} 
\DELTA m_s = 
{\rm 17.33 ^{+0.42}_{-0.21} (stat) 
\pm 0.07 (syst)}~\hbox{ps}^{-1} \,. \EQN{CDFdms} 
\end{equation} 
Both the observed significance and the expected significance of 
this signal are equal to 3.1\,$\sigma$. 
The CDF collaboration is quoting a 0.5\% probability that 
their data would fluctuate to produce, 
at any value of $\DELTA m_s$,
a fake signal as significant as the observed one,
corresponding to a 2.6\,$\sigma$ effect. 
Both D\O\ and CDF quote their $\DELTA m_s$ results assuming 
that they see the oscillation signal. 

Including all unpublished analyses%
\cite{CDF2_dms_prelim,DZERO_dms,SLD_dms_prelim} 
in the average leads to the combined amplitude spectrum 
of \Fig{amplitude} (bottom), which is dominated by the new CDF result, 
and where a consolidated signal 
is seen with a significance of 4.0\,$\sigma$. 
A preliminary world average is 
\begin{equation} 
\DELTA m_s = 17.4\, ^{+0.3}_{-0.2}~\hbox{ps}^{-1} \,. \EQN{dmsw}
\end{equation} 

The information on $|V_{ts}|$ obtained, 
in the framework of the Standard Model, from the combined amplitude spectrum,
is hampered by the hadronic uncertainty, as in the \Bd\ case. 
However, several uncertainties cancel in the frequency ratio
\begin{equation} 
\frac{\DELTA m_s}{\DELTA m_d} = \frac{m_{B_s}}{m_{B_d}}\, \xi^2
                    \left|\frac{V_{ts}}{V_{td}}\right|^2 \,,
\EQN{ratio} 
\end{equation} 
where $\xi= (f_{B_s} \sqrt{B_{B_s}})/(f_{B_d} \sqrt{B_{B_d}})
=1.210\, ^{+0.047}_{-0.035}$
is an SU(3) flavor-symmetry breaking factor 
obtained from lattice QCD calculations\cite{lattice_QCD}.
Using the averages of 
\Eqss{dmdw}{dmsw}, one can extract
\begin{equation} 
\left|\frac{V_{td}}{V_{ts}}\right| = 
{\rm 0.208 \pm 0.002 (exp)\,^{+0.008}_{-0.006} (lattice)} \,, \EQN{VtdVts}
\end{equation} 
in good agreement with (but more precise than) 
the recent result obtained by the Belle collaboration based on the 
observation of the $b\to d \gamma$ transition\cite{Belle-btodgamma}.
The CKM matrix can be constrained using experimental results 
on observables such as $\DELTA m_d$, 
$\DELTA m_s$, $|V_{ub}/V_{cb}|$, $\epsilon_K$, and $\sin(2\beta)$ 
together with theoretical inputs and unitarity 
conditions\cite{CKM_review,UTfit,CKMfitter}.
The constraint from our knowledge on the ratio $\DELTA m_s/\DELTA m_d$
is presently more effective in limiting the position of the apex of the 
CKM unitarity triangle than the one obtained from the $\DELTA m_d$ 
measurements alone, due to the reduced hadronic uncertainty in \Eq{ratio}.
We also note that the measured value of $\Delta m_s$ is consistent with the Standard 
Model prediction obtained from CKM fits
where no experimental information on $\Delta m_s$ is used, 
{\it e.g.} $21.2 \pm 3.2~{\rm ps}^{-1}$\cite{UTfit} or
$16.5\, ^{+10.5}_{-3.4}~{\rm ps}^{-1}$\cite{CKMfitter}.

Information on $\DELTA\Gamma_s$ can be obtained by studying the proper time 
distribution of untagged 
\Bs\ samples%
\cite{Hartkorn_Moser}.
In the case of an inclusive \Bs\ selection\cite{L3_DGs}, or a semileptonic (or flavour-specific)
\Bs\ decay selection\cite{DELPHI_dms_dgs,ALEPH_OPAL_CDF_Dsl_lifetime,CDF2DZERO_tauBsfs}, 
both the short- and long-lived
components are present, and the proper time distribution is a superposition 
of two exponentials with decay constants
$\Gamma_{\rm L,H} = \Gamma_s\pm \DELTA\Gamma_s/2$.
In principle, this provides sensitivity to both $\Gamma_s$ and 
$(\DELTA\Gamma_s/\Gamma_s)^2$. Ignoring $\DELTA\Gamma_s$ and fitting for 
a single exponential leads to an estimate of $\Gamma_s$ with a 
relative bias proportional to $(\DELTA\Gamma_s/\Gamma_s)^2$. 
An alternative approach, which is directly sensitive to first order in 
$\DELTA\Gamma_s/\Gamma_s$, 
is to determine the lifetime of \Bs\ candidates decaying to $CP$ 
eigenstates; measurements exist for
$\hbox{\Bs} \to J/\psi \phi$\cite{CDF1_Jpsiphi,CDF2DZERO_Jpsiphi}
and $\hbox{\Bs} \to D_s^{(*)+} D_s^{(*)-}$\cite{ALEPH_DGs}, which are 
mostly $CP$-even states\cite{Aleksan}. 
However, in the case of $\hbox{\Bs} \to J/\psi \phi$ 
this technique has now been replaced 
by more sensitive time-dependent angular analyses 
that allow the simultaneous extraction of $\DELTA\Gamma_s/\Gamma_s$ and 
the $CP$-even and $CP$-odd amplitudes\cite{CDF2DZERO_Jpsiphi_angular}.
An estimate of $\DELTA\Gamma_s/\Gamma_s$
has also been obtained directly from a measurement of the 
$\hbox{\Bs} \to D_s^{(*)+} D_s^{(*)-}$ branching ratio\cite{ALEPH_DGs}, 
under the assumption that 
these decays account for all the $CP$-even final states 
(however, no systematic uncertainty due to this assumption is given, so 
the average quoted below will not include this estimate).


Applying the combination procedure of \Ref{HFAG} 
(including the constraint from the flavour-specific lifetime measurements) on the published 
results\cite{DELPHI_dms_dgs,ALEPH_OPAL_CDF_Dsl_lifetime,CDF1_Jpsiphi,ALEPH_DGs,CDF2DZERO_Jpsiphi_angular}
yields
\begin{equation} 
\DELTA\Gamma_s/\Gamma_s = +0.31\, ^{+0.11}_{-0.13} 
~~~~\hbox{and}~~~
1/\Gamma_s = 1.398\, ^{+0.049}_{-0.050}~\hbox{ps}\,,
\EQN{DGs}
\end{equation} 
or equivalently
\begin{equation} 
1/\Gamma_{\rm L} = 1.21 \pm 0.09~\hbox{ps}
~~~~\hbox{and}~~~
1/\Gamma_{\rm H} = 1.66\, ^{+0.11}_{-0.12}~\hbox{ps}\,.
\end{equation} 
This result can be compared with the theoretical prediction 
$\DELTA\Gamma_s/\Gamma_s = +0.12\pm0.05$\cite{DGS_theory} 
within the Standard Model. 

\section*{\boldmath Average $b$-hadron mixing probability and 
$b$-hadron production fractions in $Z$ decays and at high energy}

Mixing measurements can 
significantly improve our knowledge on the fractions 
$f_u$, $f_d$, $f_s$ and $f_{\rm baryon}$, defined as 
the fractions of $B_u$, \Bd, \Bs\, and $b$-baryon 
in an unbiased sample of weakly decaying $b$~hadrons
produced in high-energy collisions. Indeed, 
time-integrated mixing analyses performed with lepton pairs 
from $b\overline{b}$ events at high energy 
measure the quantity 
\begin{equation} 
\overline{\chi} = f'_d \,\chi_d + f'_s \,\chi_s \,,
\end{equation} 
where $f'_d$ and $f'_s$ are the fractions of \Bd\ and \Bs\ hadrons 
in a sample of semileptonic $b$-hadron decays. 
Assuming that all $b$~hadrons have the same semileptonic 
decay width implies 
$f'_q = f_q/(\Gamma_q \tau_b)$ ($q=s,d$), where 
$\tau_b$ is the average $b$-hadron lifetime. 
Hence $\overline{\chi}$ measurements, together with 
the $\chi_d$ average of \Eq{chidw} and 
the very good approximation $\chi_s = 1/2$ 
(in fact $\chi_s > 0.4988$ at 95\%~CL from 
\Eqsss{chi}{dms}{DGs}),
provide constraints on the fractions $f_d$ and $f_s$.

The LEP experiments have measured 
$f_s \times {\rm BR}(B^0_s \to D_s^- \ell^+ \nu_\ell X)$\cite{LEP_fs}, 
${\rm BR}(b \to \Lambda_b^0) \times 
{\rm BR}(\Lambda_b^0 \to \Lambda_c^+\ell^- \overline\nu_\ell X)$\cite{LEP_fla},
and ${\rm BR}(b \to \Xi_b^-) \times 
{\rm BR}(\Xi_b^- \to \Xi^-\ell^-\overline\nu_\ell X)$\cite{LEP_fxi}
from partially reconstructed final states,
including a lepton, $f_{\rm baryon}$ 
from protons identified in $b$ events\cite{ALEPH-fbar}, and the 
production rate of charged $b$ hadrons\cite{DELPHI-fch}. 
The $b$-hadron fractions 
measured at CDF with electron-charm final states\cite{CDF_f} 
are at slight discrepancy with the ones measured at LEP.
Furthermore the values of $\overline{\chi}$ measured at LEP, 
$0.1259 \pm0.0042$\cite{LEPEWWG_chibar}, and at CDF, 
$0.152 \pm0.013$\cite{CDF_chibar}, 
show a 1.9\,$\sigma$ deviation with respect to each other. 
This may be a hint that the fractions at the Tevatron might be different 
from the ones in $Z$ decays.
Combining\cite{HFAG} all the available information under the constraints
$f_u = f_d$ and $f_u + f_d + f_s + f_{\rm baryon} = 1$
yields the two set of averages shown in \Table{fractions}. The second set, 
obtained using both LEP and Tevatron results, has larger errors than the first set, 
obtained using LEP results only, because we have applied scale factors 
as advocated by the PDG for the treatment of marginally consistent data. 

\begin{table}
\caption{
$\overline{\chi}$ and $b$-hadron fractions (see text).
\labt{fractions}
}
\begin{center}
\begin{tabular}{lll}
\hline\hline
                     & in $Z$ decays   & at high energy   \\
\hline
$\overline{\chi}$    & $0.1259   \pm 0.0042$ ~ & $0.1283   \pm 0.0076$ \\
$f_u = f_d$          & $0.399\,~ \pm 0.010$    & $0.398\,~ \pm 0.012$  \\
$f_s$                & $0.102\,~ \pm 0.009$    & $0.103\,~ \pm 0.014$  \\
$f_{\rm baryon}$ ~~  & $0.100\,~ \pm 0.017$    & $0.100\,~ \pm 0.020$  \\
\hline\hline
\end{tabular}
\end{center}
\end{table} 


\section*{Summary and prospects}

\BBbar\ mixing has been and still is a field of intense study.
The mass difference in the \BdBdbar\ system is now very precisely known 
(with an experimental error of $0.9\%$) but, 
despite an impressive theoretical effort, 
the hadronic uncertainty keeps limiting the precision of the 
extracted estimate of $|V_{td}|$ within the Standard Model (SM).
On the other hand measurements of $\DELTA \Gamma_d$
and of $CP$ violation in \BdBdbar\ mixing 
are consistent with zero, 
with an uncertainty still large compared to the SM predictions.
Impressive new \Bs\ results are becoming available from Run~ II 
of the Tevatron:
preliminary direct evidence for \BsBsbar\ oscillations 
has been reported, with a frequency in agreement with the SM.
New time-dependent angular analyses of 
$\hbox{\Bs} \to J/\psi \phi$ decays at CDF and D\O\
have improved our knowledge of 
$\DELTA \Gamma_s/\Gamma_s$ to an absolute uncertainty 
of $\sim 10\%$, of the same size as the central value
of the SM prediction. The data clearly prefer 
$\Gamma_{\rm L} > \Gamma_{\rm H}$ as predicted in the SM.

Improved results on \BsBsbar\ mixing are still to come from the Tevatron, 
with very promising prospects in the next couple 
of years, both for $\DELTA m_s$ and $\DELTA \Gamma_s$. 
With a few $\hbox{fb}^{-1}$ of data, 
the CDF and D\O~collaborations will have the potential to confirm
their $\DELTA m_s$ signals and make $>5\,\sigma$ observations 
of \Bs\ oscillations. Further studies with $\hbox{\Bs} \to J/\psi \phi$ decays
will not only improve on $\DELTA \Gamma_s$, but perhaps also allow 
a very first investigation of the $CP$-violating phase $\phi_s$ induced 
by \BsBsbar\ mixing, 
about which nothing is known experimentally at present. However, the SM value 
of $\phi_s$ is very small ($\phi_s=-2\beta_s$ where 
$\beta_s \equiv \arg(-V_{ts}^{}V_{tb}^*/(V_{cs}^{}V_{cb}^*))$ 
is about one degree), 
and a full search for new physics effects in this observable will require 
much larger statistics. These will become available at CERN's Large Hadron Collider 
scheduled to start operation in 2007, where the 
%
LHCb 
%
collaboration expects to be able to measure
$\phi_s$ down to the SM value after several years of operations\cite{LHCb}.



$B$ mixing may not have delivered all its secrets yet, because 
it is one of the phenomena where new physics might still reveal itself
(although a dominant contribution is becoming unlikely).
Theoretical calculations in lattice QCD have become more reliable, and 
further progress in reducing hadronic uncertainties is expected. 
In the long term, a stringent check of the consistency, within the 
SM, of the \Bd\ and \Bs\ mixing amplitudes (magnitudes and phases) 
with all other measured flavour-physics observables
(including $CP$ asymmetries in $B$ decays) will be possible, 
leading to further limits on new physics or, better, 
new physics discovery.

\end{document}